\documentclass[sn-mathphys]{sn-jnl}% Math and Physical Sciences Reference Style
\jyear{2022}%

\theoremstyle{thmstyleone}%
\theoremstyle{thmstyletwo}%

\theoremstyle{thmstylethree}%

\begin{document}

\title[First Steps of Planet Formation Around VLMS and BDs]{First Steps of Planet Formation Around Very Low Mass Stars and Brown Dwarfs}

\author[1,2]{\fnm{Paola} \sur{Pinilla}}\email{pinilla@mpia.de}

\affil[1]{\orgdiv{Mullard Space Science Laboratory}, \orgname{University College London}, \orgaddress{\street{Holmbury St Mary}, \city{Dorking}, \country{UK}}}

\affil[2]{\orgdiv{Planet and Star Formation Department}, \orgname{Max-Planck-Institut f\"{u}r Astronomie}, \orgaddress{\street{K\"{o}nigstuhl 17}, \city{Heidelberg},  \country{Germany}}}


\abstract{Brown dwarfs and very low mass stars are a significant fraction of stars in our galaxy and they are interesting laboratories to investigate planet formation in extreme conditions of low temperature and densities. In addition, the dust radial drift of particles is expected to be a more difficult barrier to overcome during the first steps of planet formation in these disks. ALMA high-angular resolution observations of few protoplanetary disks around BDs and VLMS have shown substructures as in the disks around Sun-like stars. Such observations suggests that giant planets embedded in the disks are the most likely origin of the observed substructures. However, this type of planets represent less than 2\% of the confirmed exoplanets so far around all stars, and they are difficult to form by different core accretion models (either pebble or planetesimal accretion). Dedicated deep observations of disks around BDs and VLMS with ALMA and JWST  will provide significant progress on understanding the main properties of these objects (e.g., disk size and mass), which is crucial for determining the physical mechanisms that rule the evolution of these disks and the effect on the potential planets that may form in these environments. }

\maketitle

\section{Introduction}\label{sec1}
\begin{figure}[h]
\centering
\includegraphics[width=0.7\textwidth]{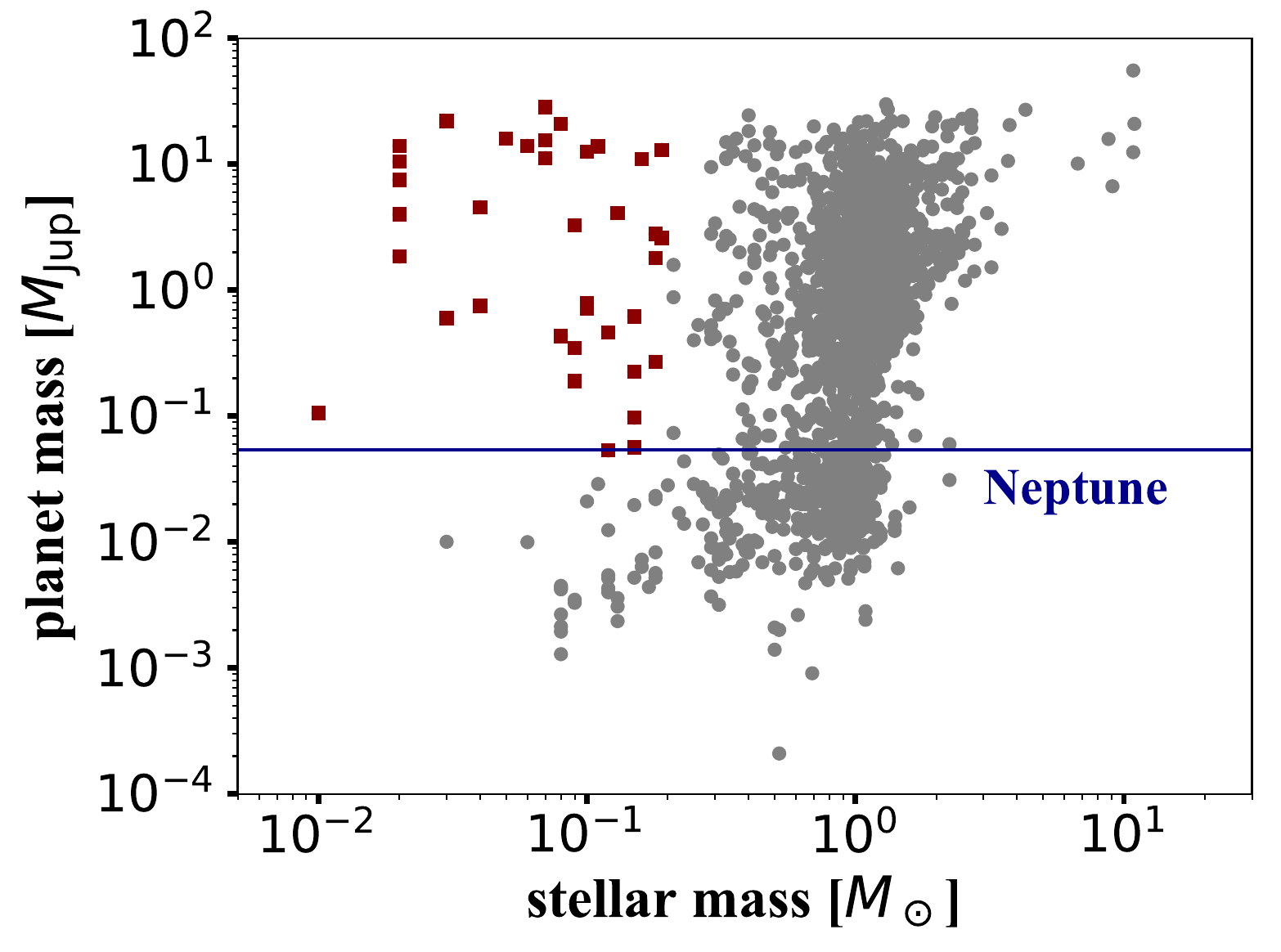}
\caption{Observed giant exoplanets ($M>1$\,$M_{\rm{Nep}}$) around BDs and VLMS ($M<0.2\,M_\odot$, red squares) compared to confirmed exoplanets around all type of stars. This population is less than 2\% of the detected exoplanets so far. This percentage is not observational-bias-corrected}. Data retrieved on 2022.06.01 from \url{https://exoplanetarchive.ipac.caltech.edu }.\label{exoplanets}
\end{figure}

Brown Dwarfs (BDs) are substellar objects in the mass range between 13-80 Jupiter masses approximately. They are unable to sustain hydrogen fusion in their cores, and they represent the link between the lowest mass stars (M-dwarfs) and (exo-)planets. BDs together with very low-mass M-dwarf stars of masses $\sim0.1-0.2\,M_\odot$ (referred as very low mass stars in this paper - VLMS)  represent a significant fraction of stars in our galaxy \citep[15-20\%,][]{Chabrier2003, Kirkpatrick2012, Muzic2017}. They are interesting laboratories to investigate planet formation because the frequency of Earth-sized planets in habitable zones appears to be higher around M-dwarfs \citep{mulders2015,hardegree2019, sabotta2021}, such as the exciting cases of TRAPPIST\,1 system \citep{gillon2016} and Proxima Centauri\,b \citep{anglada2016}. Contrary, the frequency of giant planets around M-dwarfs seems to be low \citep{johnson2010,Bonfils2013, obermeier2016, ghezzi2018}.  Figure~\ref{exoplanets} shows the observed giant exoplanets ($M>1$\,$M_{\rm{Nep}}$) around BDs and VLMS ($M<0.2\,M_\odot$) compared to confirmed exoplanets around all type of stars. The population of giant planets around BDs and VLMS is less than 2\% of the detected exoplanets so far.

These trends of exoplanet demographics may be reflected in the properties of the birth-sites of planets (protoplanetary disks) around BDs and VLMS. 
Observations of young BDs and VLMS show near-infrared excess emission \cite{apai2005, luhman2006, riaz2006, scholz2006, scholz2007, daemgen2016} that reveals the existence of protoplanetary disks around these objects. Several of these disks have been detected with (sub-)millimeter observations in nearby star forming regions \cite{klein2003, scholz2006b, joergens2012, ricci2012, ricci2014, ricci2017, mohanty2013, testi2016, vanderplas2016, pinilla2017, wardduong2018, sanchis2020, rilinger2021}, with typical fluxes lower than few mJy at 1\,mm, implying disks dust masses of a few Earth masses and lower.  

These observations of BDs and VLMS disks at sub-mm/mm wavelengths suggest that even in these very low-mass and cold conditions, $\mu$m-sized dust particles grow to mm-sizes and these pebbles remain in the disk for 1-5\,Myr timescales \cite{ricci2012, ricci2017, pinilla2017}.  The initial growth of particles in protoplanetary disks is governed by the interaction with the gas. Because of the expected sub-Keplerian motion of the gas, the dust particles lose angular momentum and drift inward very fast ($\lesssim 1000$ yr). Theoretical models predict that the radial-drift barrier is a more difficult problem to overcome for mm-dust particles in disks around VLMS, in particular BDs, compared to Sun-like stars.

Millimeter observations of BD and VLMS disks have challenged current dust evolution models of protoplanetary disks. Only under extreme conditions where the radial drift is highly reduced in the entire disk can dust evolution models explain the current millimeter observations of BD and VLMS disks. To reduce the high radial drift, it has been suggested that there must be strong pressure bumps that are radially distributed in the disk \cite{pinilla2013}. The presence of such pressure bumps leads to  bright ring-like structures observable at optical, near-infrared, and (sub-)millimeter wavelengths, as have recently been observed in disks around Sun-like and Herbig stars \cite[e.g.,][]{alma2015, andrews2018, andrews2020, long2018, cieza2021}. A possible origin for such pressure bumps is embedded planet(s) in the disk. 

For disks around BDs and VLMS, a planet with at least a Saturn mass is required to open a gap in the gas surface density and trap millimeter-sized particles \cite{pinilla2017, sinclair2020}. This is challenging because Saturn-mass planets are difficult to form in models of disks around BDs and VLMS, either by pebble or planetesimal accretion \citep{liu2020, miguel2020, burn2021}. Alternatively, giant planets around M-dwarfs may form when the disk is gravitational unstable \citep[][]{mercer2020}. 

In this paper, we present state-of-the art observations of the dust continuum emission of disks around BDs and VLMS that have been observed with ALMA and show substructures (Sect.~\ref{ALMA_VLMS}). These observations are a motivation to revise the dust radial drift barrier in these disks when stellar evolution is taken into account in the first million of years (Sect.~\ref{radial_drift}). To stop the fast radial drift of dust particles in disks around BDs and VLMS, we study the type of planet that is required to open a gap in a disk around a 0.1\,$M_\odot$ with typical stellar parameters at 1\,Myr of evolution. We perform hydrodynamical simulations of the gas and dust evolution in a disk around a typical VLMS and show the type of substructures that are expected when observed  at high angular resolution at different wavelengths (Sect.~\ref{hydro_models}). We summarize the results in Sect.~\ref{summary}.

\begin{table}[h]
\begin{center}
\begin{minipage}{\textwidth}
\caption{Disks around BDs and VLMS (Spectral type higher than M4.5) that show substructures from available ALMA observations }\label{targets_ALMA}%
\begin{tabular*}{\textwidth}{@{\extracolsep{\fill}}lccccccc@{\extracolsep{\fill}}}
\toprule%
Source & SpT  & $M_\star[M_\odot]$ & $\nu$[GHz] &$F_{\nu}$[mJy] & Region & $d$[pc]&Reference\\
\midrule
Iso-Oph196&M5 &0.20&230.0&98.0&Ophiuchus&137.0&\cite{cieza2021}\\
Iso-Oph2B & ? & 0.08&230.0&1.3&Ophiuchus&144.0&\cite{gonzalez2020}\\
ZZTauIRS & M5&0.20&339.0&273.9&Taurus&130.7&\cite{hashimoto2021}\\
CIDA1 & M4.5 & 0.19&334.5&31.0&Taurus&137.5&\cite{pinilla2021}\\
MHO6 & M5 & 0.17& 334.6&48.4&Taurus&141.9& \cite{kurtovic2021}\\
J04334465&M5.2&0.15&334.7&37.9&Taurus&173.3& \cite{kurtovic2021}\\
J04295422&M4&0.20&339.0&21.4&Taurus&140.0&[*]\\
J04394748&M7&0.10&339.0&3.3&Taurus&141.1&[*]\\
J16070854&M5&0.17&335.0&92.0&Lupus&200.0&\cite{vandermarel2018}\\
Sz100&M5&0.18&335.0&54.9&Lupus&200.0&\cite{ansdell2018}\\
Sz76&M4&0.23&230.0&3.4&Lupus&160.0&\cite{vanderMarel2022}\\
Sz103&M4&0.23&230.0&4.5&Lupus&160.0&\cite{vanderMarel2022}\\
J16081497&M5.5&0.10&230.0&3.4&Lupus&159.0&\cite{vanderMarel2022}\\
J16090141&M4&0.15&230.0&5.3&Lupus&164.0&\cite{vanderMarel2022}\\
J16070384&M4.5&0.16&230.0&1.1&Lupus&159.0&\cite{vanderMarel2022}\\
\botrule
\end{tabular*}
\footnotetext{Columns: target name, spectral type (SpT), stellar mass, frequency of the observations, total flux, target's region, distance to the source, and reference.}
\footnotetext[*]{Data from the ALMA archive from project ID 2016.1.01511.S.}
\end{minipage}
\end{center}
\end{table}

\begin{figure}%
\centering
\includegraphics[width=0.9\textwidth]{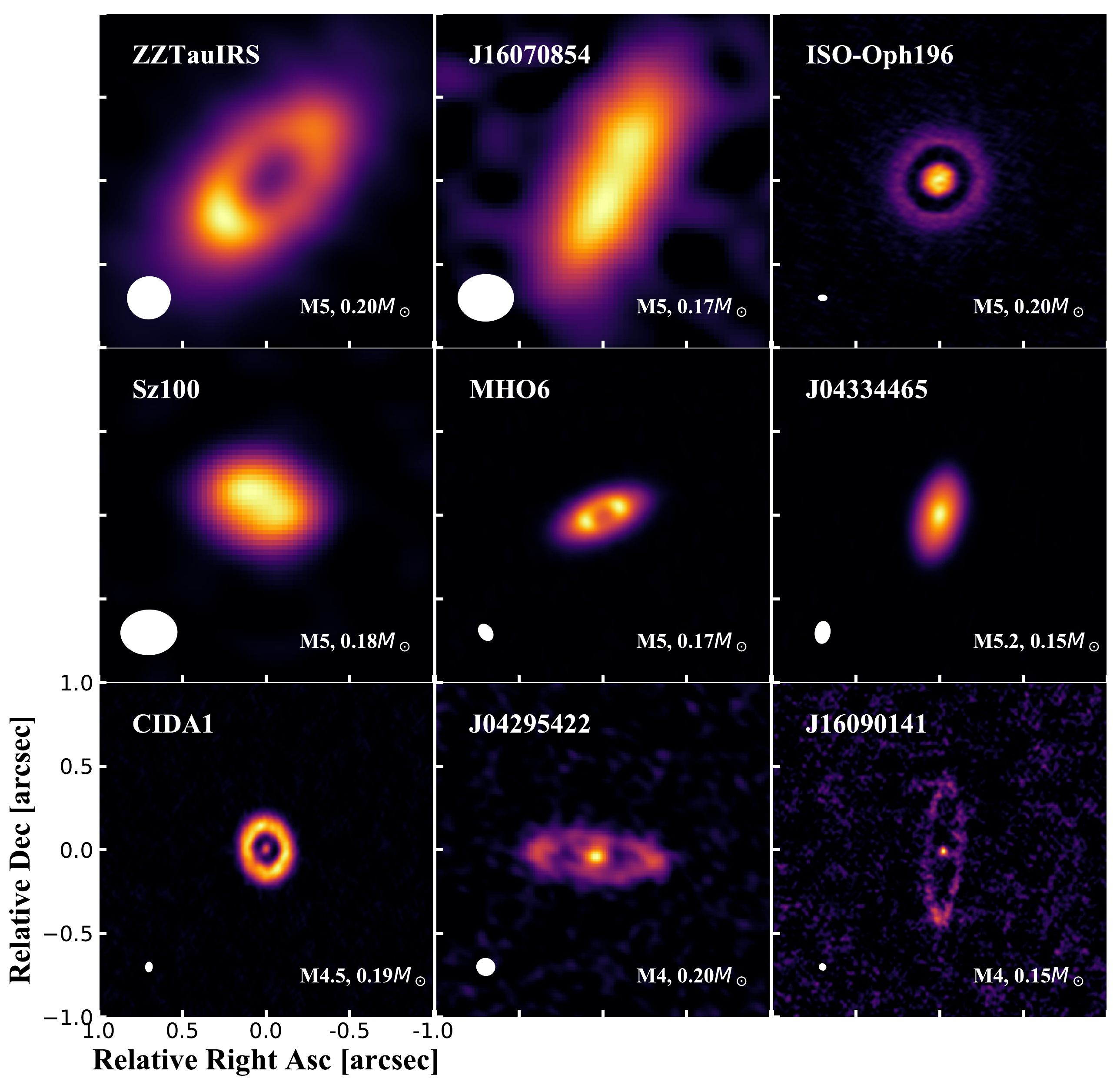}
\includegraphics[width=0.9\textwidth]{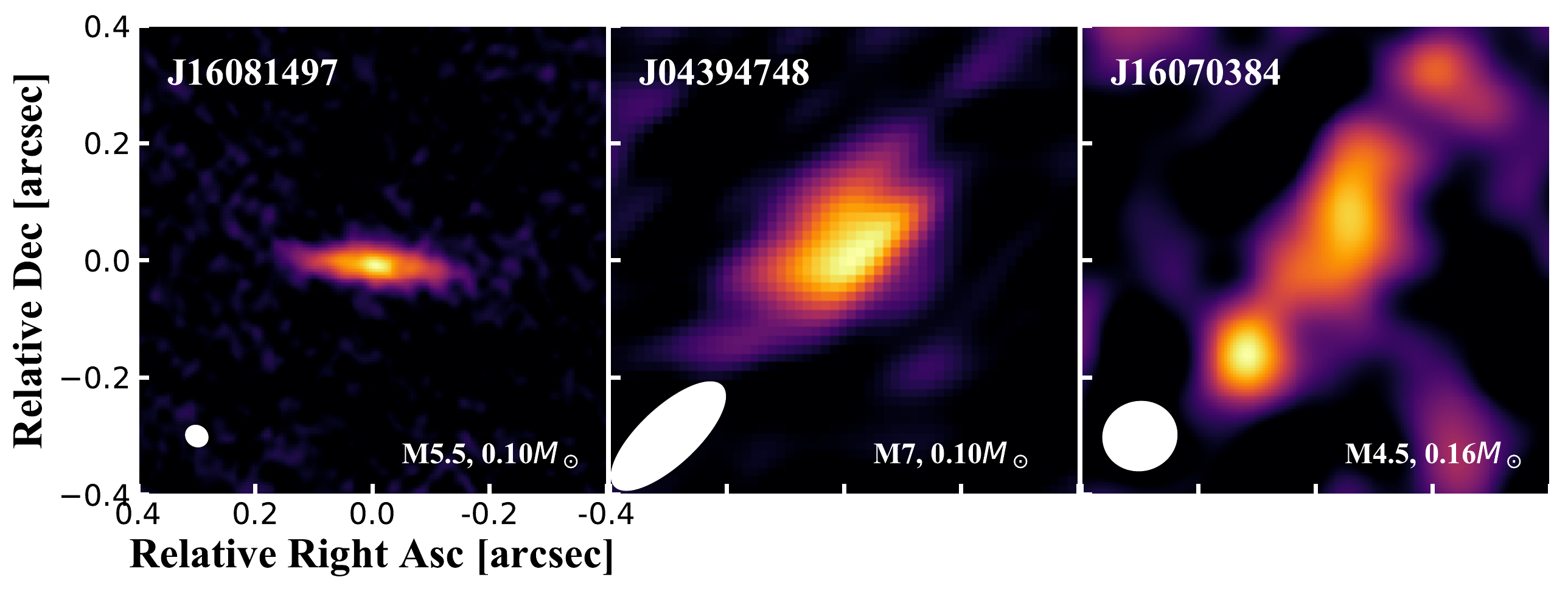}
\includegraphics[width=0.9\textwidth]{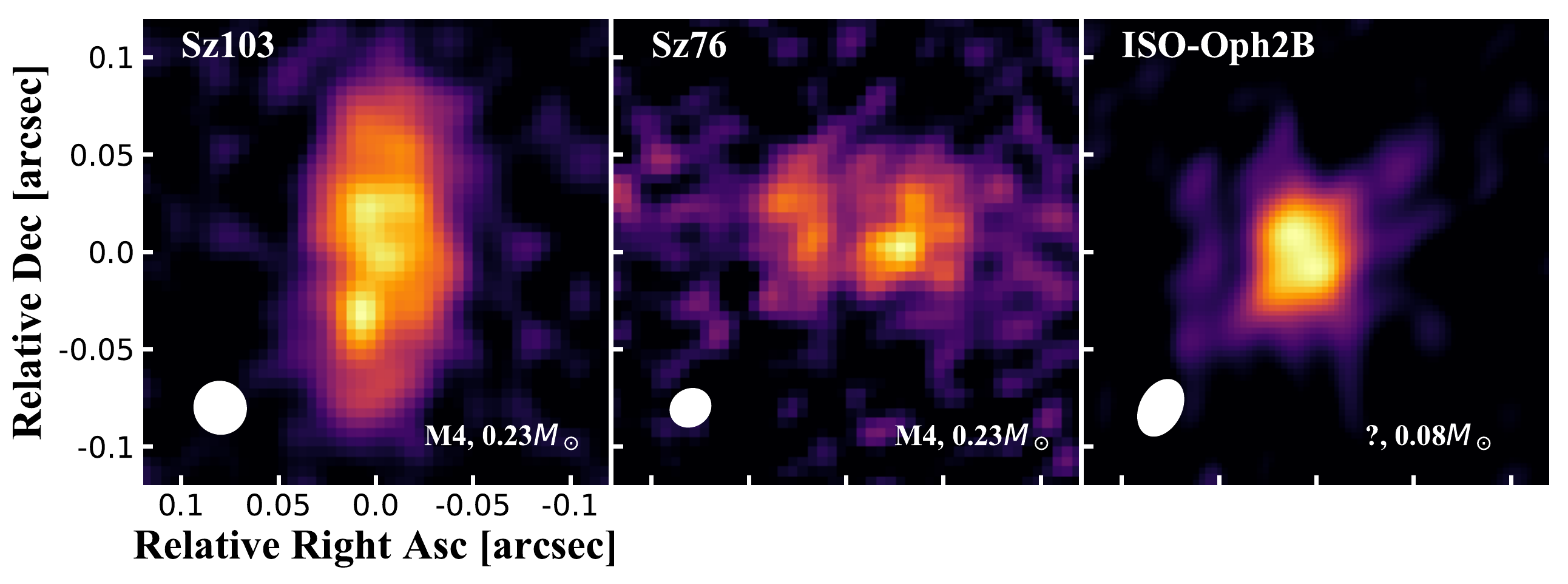}
\caption{Disks around BDs and VLMS (Spectral type higher than M4.5, Table~\ref{targets_ALMA}) that show substructures from available ALMA observations. Note that the scale of the images changes between the three set of panels from the top to the bottom.}\label{VLMS_ALMA}
\end{figure}

\section{ALMA observations of disks around BDs and VLMS} \label{ALMA_VLMS}

\begin{figure}%
\centering
\includegraphics[width=0.7\textwidth]{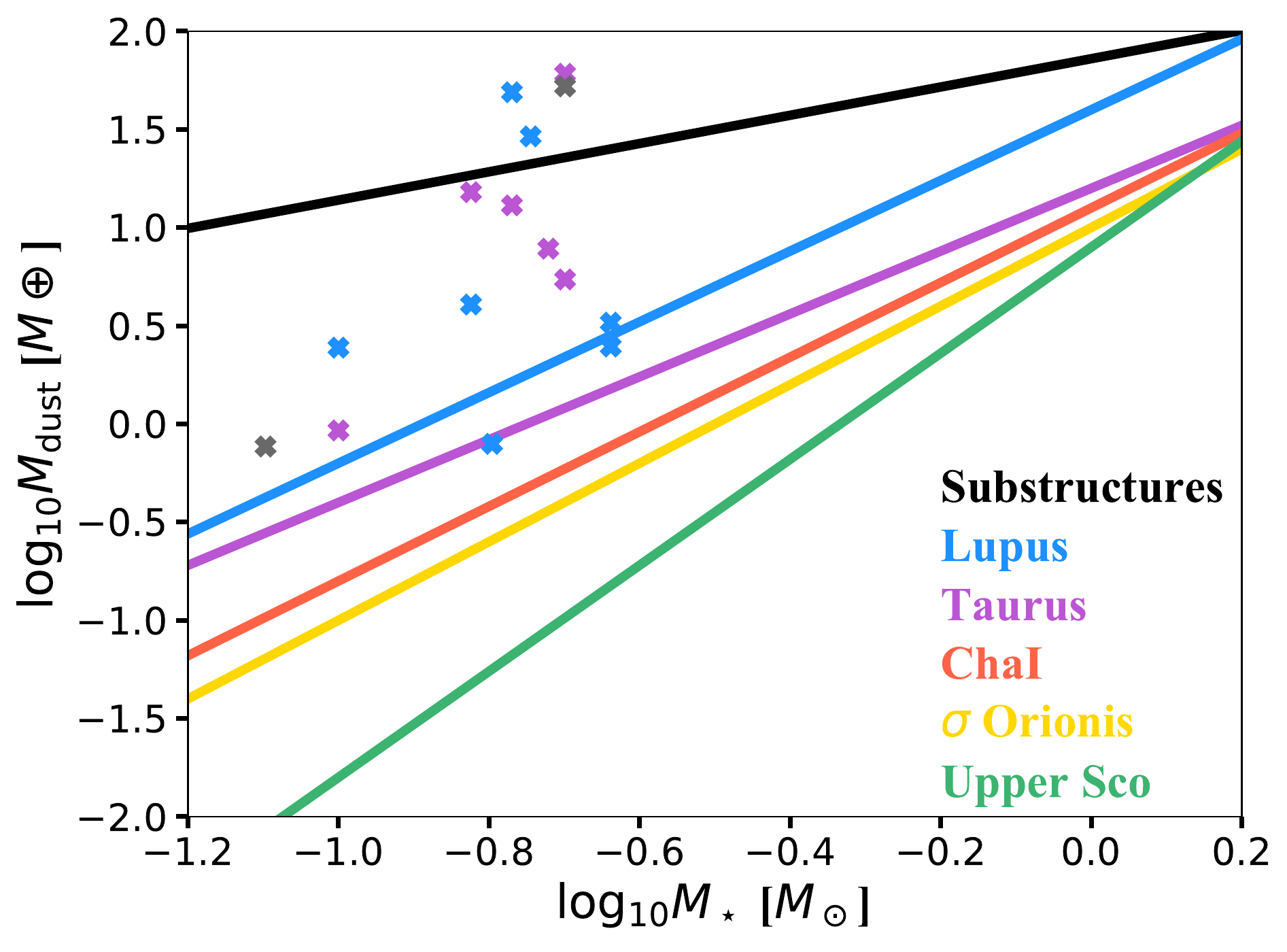}
\caption{$M_{\rm{dust}}-M_\star$ relation in different star-forming regions (colours, from the youngest regions 1-3\,Myr, Lupus \& Taurus; to the oldest 8-10\,Myr, Upper Sco). The values for the slope and intercept for the different star forming regions are taken from \cite{pascucci2016} and \cite{ansdell2017}. The black line corresponds to the observed relation to disks with substructures \citep{pinilla2020}. The cross-symbols correspond to the disks in Table~\ref{ALMA_VLMS} with the corresponding colours from the star-forming region that they belong. The grey crosses are for the two disks in Ophiuchus} \label{Mdisk_Mstar}
\end{figure}

From sub-mm/mm surveys of young stars (1-10\,Myr old) with ALMA, the dust masses inferred from the dust continuum emission of their disks shows a declining trend with the central object mass (Fig.~\ref{Mdisk_Mstar}). The scaling laws between dust mass and stellar/substellar mass inferred in different nearby star forming regions follow an approximately a power-law relation that steepens in time, although derived power-law slopes depend strongly upon choices of stellar evolutionary model and luminosity and dust temperature relation \citep{pascucci2016, testi2016, wardduong2018}. The observed $M_{\rm{dust}}-M_\star$ relation in different star forming regions can be reproduced when radial drift of particles is more efficient around VLMS as predicted from the dust evolution models \citep{pinilla2013, pinilla2020}.

The $M_{\rm{dust}}-M_\star$ relation seems to be flatter for disks with substructures, implying that these disks remain massive in dust independent of the mass of the stellar host. The flatness of the $M_{\rm{dust}}-M_\star$ relation for disk with substructures can be explained by the presence of multiple pressure traps formed by giant planets, when planetesimal formation is hindered inside these pressure bumps \citep{pinilla2020}.

Currently, the flatness of the $M_{\rm{dust}}-M_\star$ relation for disks with substructures may be a bias because high angular resolution observations with ALMA have been mostly focused on bright and large disks around T-Tauri and Herbig stars. Only a handful of disks populate the $M_{\rm{dust}}-M_\star$ relation in the stellar regime of BDs and VLMS, and the $M_{\rm{dust}}-M_\star$ relations in this stellar mass region are dominated by the fluxes of T-Tauri and Herbig disks. High angular resolution of disks around BDs and VLMS are challenging due to their faint emission and their small size. 

However, recent efforts with ALMA have been done to resolve the disks and potentially detect substructures around BDs and VLMS.  The work from \cite{kurtovic2021} presented a small sample of 6 disks around VLMS in Taurus, where 50\% of the sample revealed
substructures at a resolution of 0.1” (these are included in Table~\ref{targets_ALMA} and Fig.~\ref{ALMA_VLMS}). The sample in \cite{kurtovic2021} was selected to target the brightest disks in the very-low mass regime of the central object, and hence it was likely to find substructures as in the case of T-Tauri and Herbig disks. Additional observations have been done at high angular resolution that include BDs and VLMS revealing substructures \citep{gonzalez2020,hashimoto2021, vanderMarel2022, cieza2021, pinilla2021} mainly in Taurus and Lupus star-forming regions (Table~\ref{targets_ALMA} and Fig.~\ref{ALMA_VLMS}).

Nine of the disks in Fig.~\ref{ALMA_VLMS} are large and comparable in size to disks around T-Tauri stars (top panels in Fig.~\ref{ALMA_VLMS}). These nine disks mainly show large inner gaps, typically seen in the so-called transition disks. Four  of them show a bright emission in the inner disk. Only one of them show a clear evidence of an azimuthal asymmetry (ZZTauIRS, \cite{hashimoto2021}), a characteristic that has been observed in few disks around T-Tauri and Herbig stars \citep[e.g.,][]{marel2021}. The other six disks are very compact, and their substructures are more blurred. Currently, it is unknown if most of the compact disks are small because of the lack of substructures or if they have substructures as the large disks around T-Tauri and Herbig stars. These tentative substructures in the bottom panels of Fig.~\ref{ALMA_VLMS} suggest that compact disks also host substructures in particular in BDs and VLMS disks, and higher angular resolution and sensitivity observations are required to reveal the nature of these potential substructures. 

These total of fifteen disks around BDs and VLMS that have substructures are placed in $M_{\rm{dust}}-M_\star$ relation in Fig.~\ref{Mdisk_Mstar}. Half of this sample is closer to the flat relation obtained for disks with substructures around T-Tauri and Herbig disks, while the other half is closer to the obtained relations from large samples where substructures are not yet detected in the majority of disks. This demonstrates the need for surveys  dedicated to the low mass regime of the central objects to characterize the main properties of these disks (e.g., dust mass and size) and put them in context of the $M_{\rm{dust}}-M_\star$ relation and the disk size luminosity relation $R_{\rm{dust}}-L_{\rm{mm}}$ that are found mostly from disks around T-Tauri and Herbig stars \citep[e.g.,][]{tripathi2017, hendler2020}. Understanding the origin of these observed relations help to constrain the main physical processes of the disk evolution \citep[e.g.][]{rosotti2019, toci2021, zormpas2022}.\\

Detailed analysis of the origin of the observed substructures in disks around BDs and VLMS exclude ice-lines of the main volatiles in the disk as potential origin of the observed rings \citep[e.g.,][]{kurtovic2021, pinilla2021}. This is because the expected location of different ice-lines (e.g., H$_2$O and CO) is closer-in than the actual location of the rings. Magneto-hydrodynamical simulations predict a region of low disk turbulence (the so-called dead zone), and at the edge of that region rings may form \citep[e.g.,][]{flock2015}. Because these models have been focused on stars like our Sun, it is not clear if dead zones are expected in disks around BDs and VLMS. A parametric study of a model that captures the essence of the MRI-driven accretion presented in \citep{delage2022} suggests that dead zones in disks around VLMS may have a dead zone with a radial extension up to 3-4\,au, which is also much closer than the observed rings. The inner edge of a dead zone around  BDs and VLMS is expected in the first au, which can help for the formation of close-in small Kepler planets around these low-mass stars, or alternatively help for stopping the inward migration of planets that form further-out. Another possibility for the formation of the observed ALMA rings is planet-disk interaction, but the planet masses required to explain the observations are as high as Saturn and Jupiter mass \citep[e.g.][]{pinilla2017, curone2022}.

\section{Radial drift around BDs and VLMS}\label{radial_drift}

The very first step of planet formation includes the growth from micron-sized particles from the interstellar medium to pebbles. One of the main problems in these first steps is the high radial drift of dust particles \citep[e.g.][]{Weidenschilling1977}. It has been demonstrated that these radial drift velocities are higher for disks around BDs and VLMS \citep{pinilla2013, zhu2018}. 

\begin{figure}
\centering
\centering
    \tabcolsep=0.05cm 
    \begin{tabular}{cc}   
    	\includegraphics[width=0.5\columnwidth]{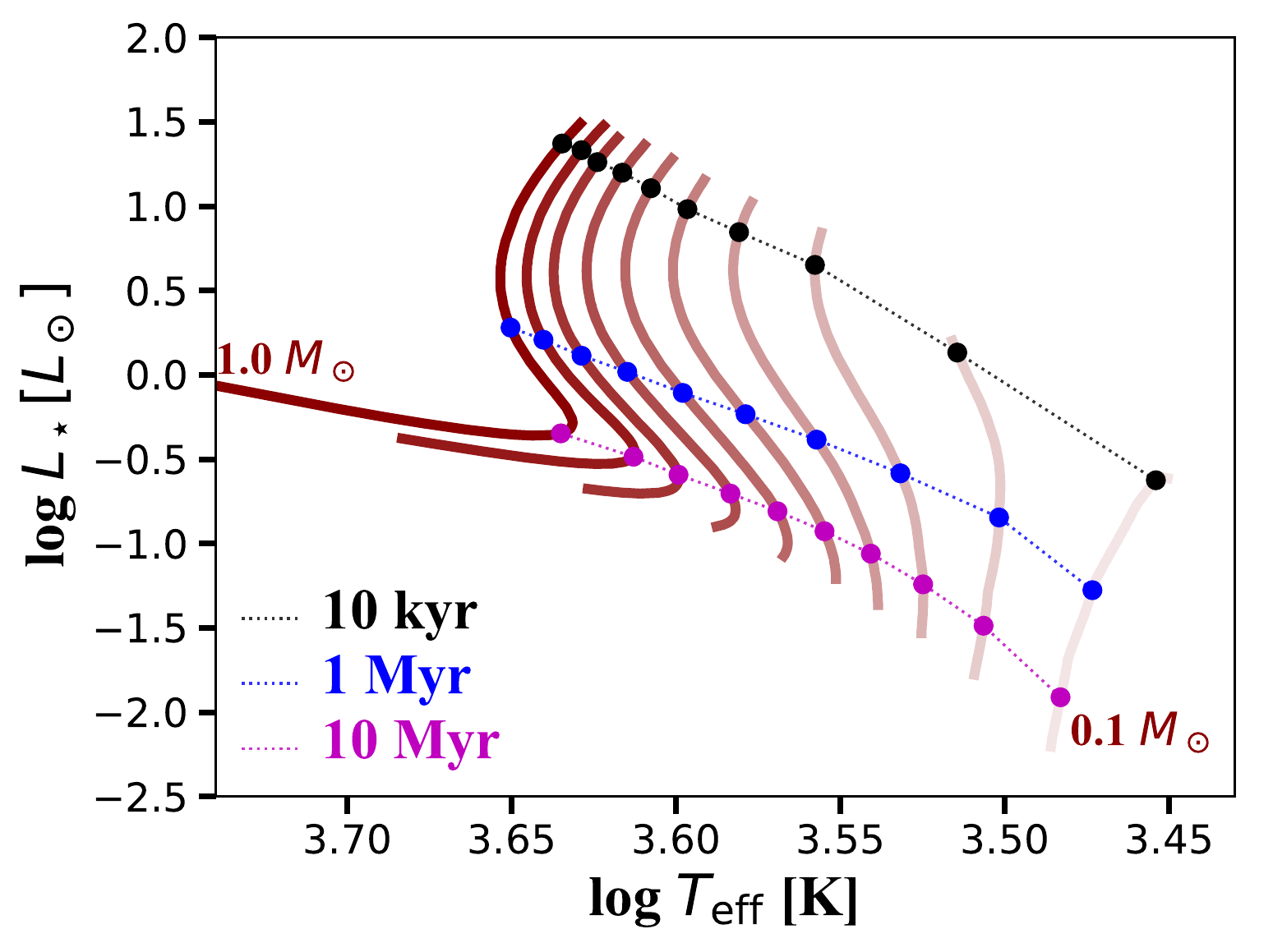}&
    	\includegraphics[width=0.5\columnwidth]{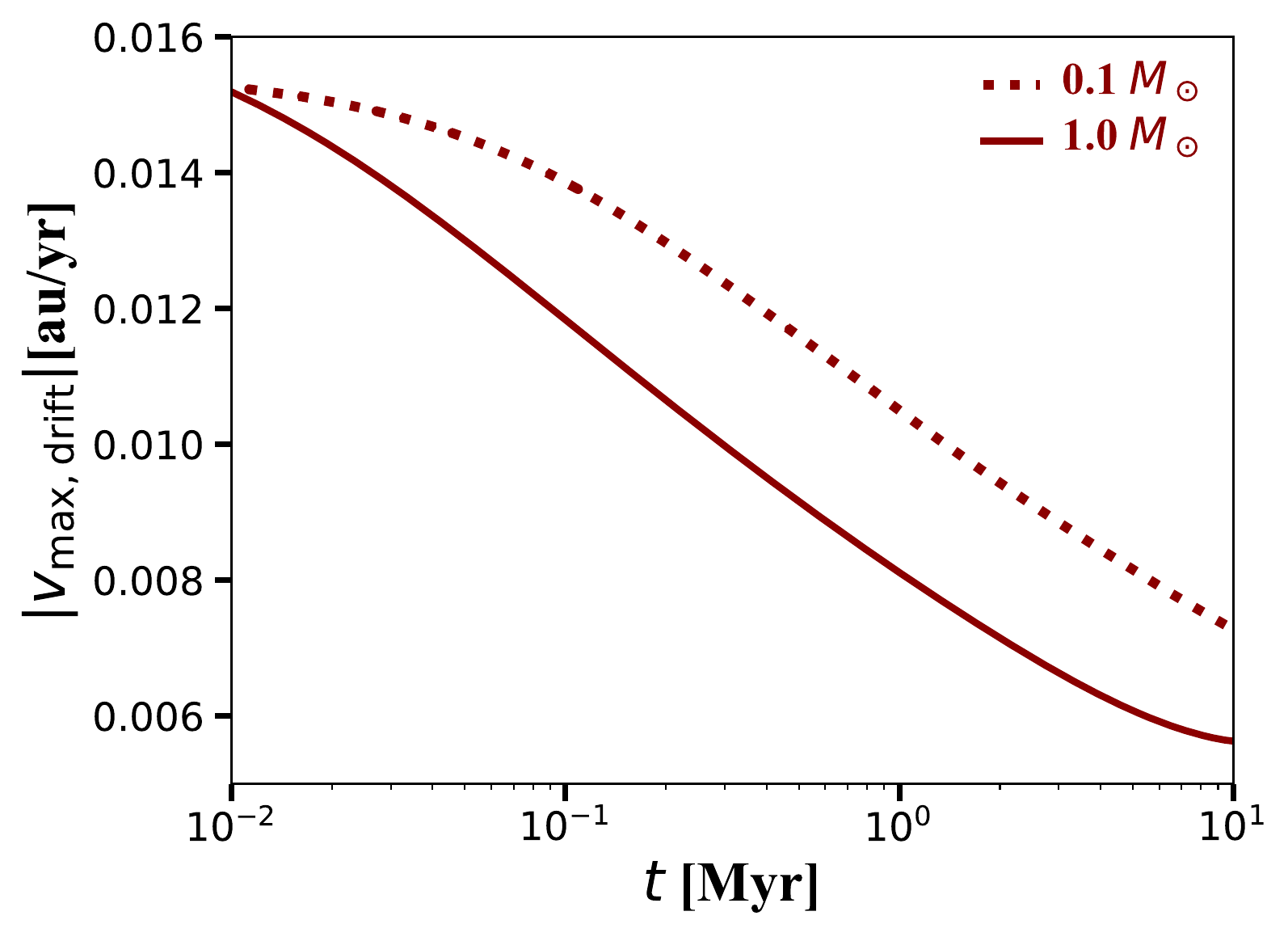}
    \end{tabular}
\caption{Left panel: Evolutionary tracks from \cite{dotter2008} up to 30\,Myr of 0.1-1\,$M_\odot$ stars in the H-R diagram, with three different isochronic lines (from light to dark red every 0.1\,$M_\odot$). Right panel: Absolute value of the maximum drift velocity as a function of time of dust particles around a 0.1\,$M_\odot$ and 1\,$M_\odot$.}\label{PMS}
\end{figure}

This is because the  radial drift velocities of dust particles depend on the stellar mass and luminosity  as $v_{\rm{drift}}\propto L_\star^{1/4}/\sqrt{M_\star}$, where the dependency with stellar mass dominates for the low mass regime, while the stellar luminosity dominates for intermediate-mass stars  $M_\star>2.5M_\odot$ \cite{pinilla2022}. Figure~\ref{PMS} shows the stellar evolution up to 30\,Myr of 0.1-1\,$M_\odot$ stars in the H-R diagram (left panel) and the absolute value of the maximum drift velocity as a function of time of dust particles around a 0.1\,$M_\odot$ and 1\,$M_\odot$. During the typical ages of protoplanetary disks (up to $\sim10$\,Myr), the dust radial drift velocities of dust particles are higher around a 0.1\,$M_\odot$ than around 1\,$M_\odot$. We note that this difference is higher for BDs, but due to the lack of evolutionary tracks for objects less massive than 0.1\,$M_\odot$ we cannot include BDs in Fig~\ref{PMS}.

To quantify the effect of radial drift in disks around VLMS, we perform dust evolution models using the publicly available code \texttt{Dustpy}\footnote{
\texttt{Dustpy} is available on \href{https://github.com/stammler/DustPy}{github.com/stammler/DustPy}} (version 0.59) \citep{stammler2022}. This code simultaneously calculates the transport of the grains and the dust growth. For our purpose, we perform simulations for disks around a 0.1 and 1\,$M_\odot$ while the stellar luminosity evolves with time as shown in the left panel of Fig.~\ref{PMS}. We assume that $M_{\rm{disk}}=0.05M_\star$ and a disk radial extension from 5 to 100\,au in both cases. The disk viscosity which also controls the dust diffusion and turbulence (radially and vertically) is taken to be $\alpha=10^{-3}$. We consider that the fragmentation velocity ($v_{\rm{frag}}$) is 10~m\,s$^{-1}$. The actual value of these velocities is currently debatable because laboratory experiments are limited to temperatures that are higher than the typical disk miplane temperatures \citep{musiolik2019, Pillich2021}.

\begin{figure}
\centering
\centering
    	\includegraphics[width=0.8\columnwidth]{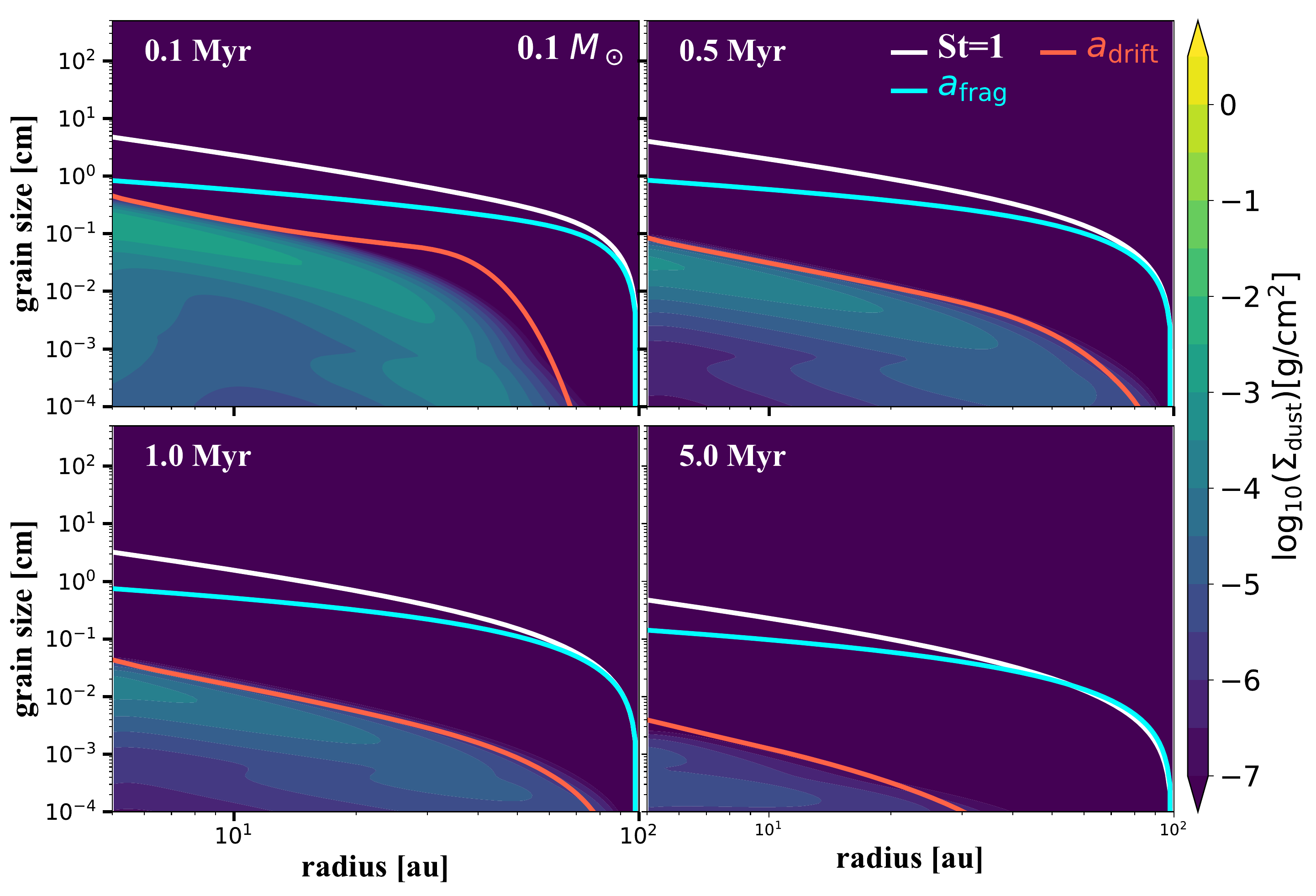}\\
    	\includegraphics[width=0.8\columnwidth]{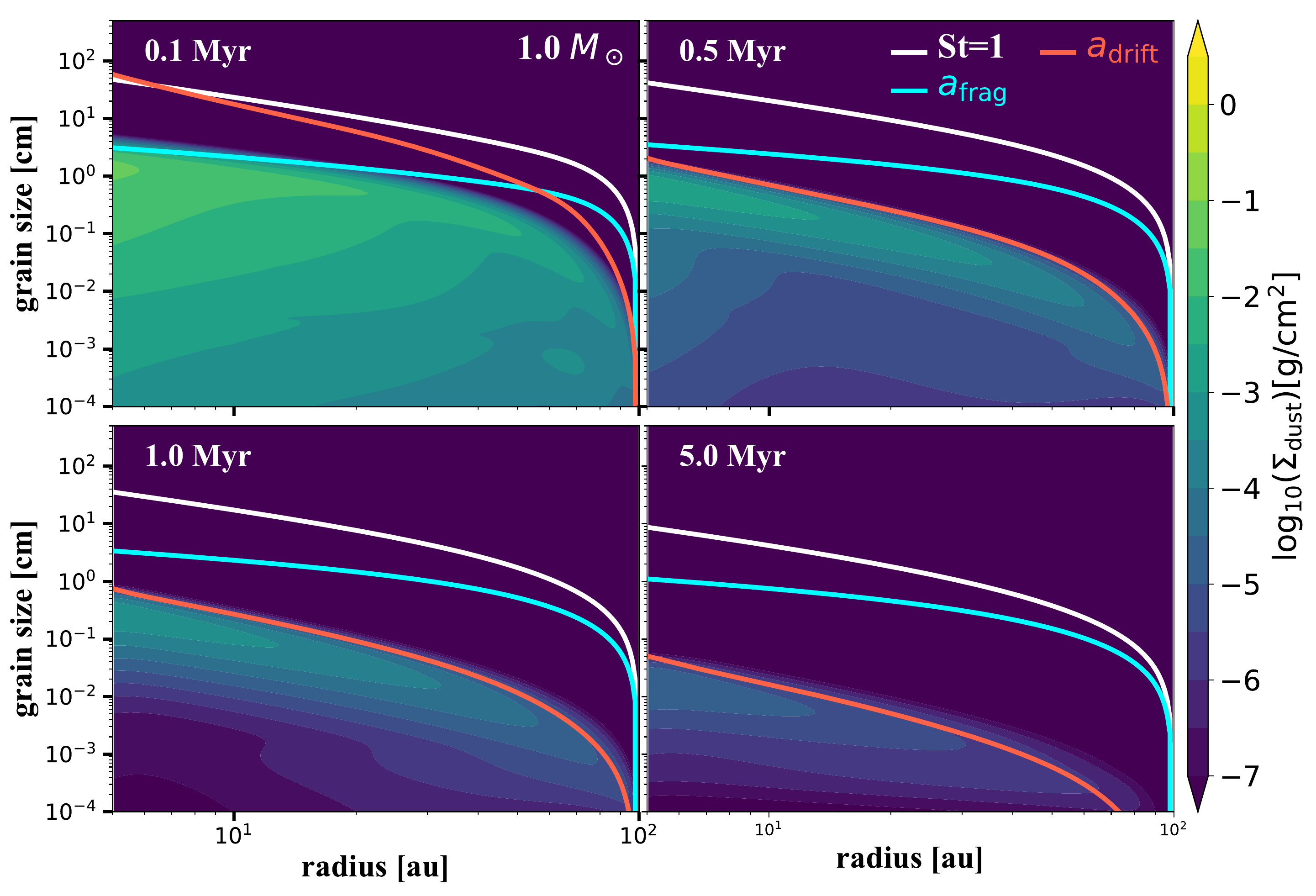}
\caption{Dust density distribution at four different times of evolution (0.1, 0.5, 1.0 and 5.0\,Myr) as a function of distance from the star and grain size for models of disks around 0.1\,$M_\odot$ (upper panels) and 1\,$M_\odot$ (bottom panels), and considering stellar evolution. The solid white line represents a Stokes number of unity (which is proportional to the gas surface density), while cyan and red show the fragmentation and drift barrier, respectively.}\label{dust_evo_variablestar}
\end{figure}

The dust density distribution at four different times of evolution of these simulations are shown in Fig.~\ref{dust_evo_variablestar}. Each panel in this figure includes the following lines: the Stokes number of unity, the fragmentation and the drift barriers. 

The dimensionless Stokes number quantifies the coupling of dust particles to the gas and in the Epstein drag regime  in the midplane of the disk is defined as \citep{birnstiel2010}

\begin{equation}
\textrm{St}=  \frac{a\rho_s}{\Sigma_g}\frac{\pi}{2},
\label{eq:stokes}
\end{equation}

\noindent where $a$ is the grain size, $\rho_s$ is the intrinsic volume density of the particles, which we set to 1.67\,g\,cm$^{-3}$, and $\Sigma_g$ is the gas surface density. Therefore, St=1 represents also the shape of the gas surface density.

The fragmentation barrier is set assuming that the relative velocities of particles are dominated by turbulence and that particles stick in collisions below a given fragmentation speed $v_{\rm{frag}}$, and it is defined as \citep{birnstiel2010}

\begin{equation}
	a_{\mathrm{frag}}=\frac{2}{3\pi}\frac{\Sigma_g}{\rho_s \alpha}\frac{v_{\rm{frag}}^2}{c_s^2}.
  \label{eq:afrag}
\end{equation}

The drift barrier occurs when growth is limited by radial drift and is given by \citep{birnstiel2010}

\begin{equation}
	a_{\mathrm{drift}}=\frac{2 \Sigma_d}{\pi\rho_s}\frac{v_K^2}{c_s^2}\left \vert \frac{d \ln P}{d\ln r} \right \vert^{-1},
  \label{eq:adrift}
\end{equation}

\begin{figure}
\centering
\centering
    \tabcolsep=0.05cm 
    \begin{tabular}{cc}   
    	\includegraphics[width=0.5\columnwidth]{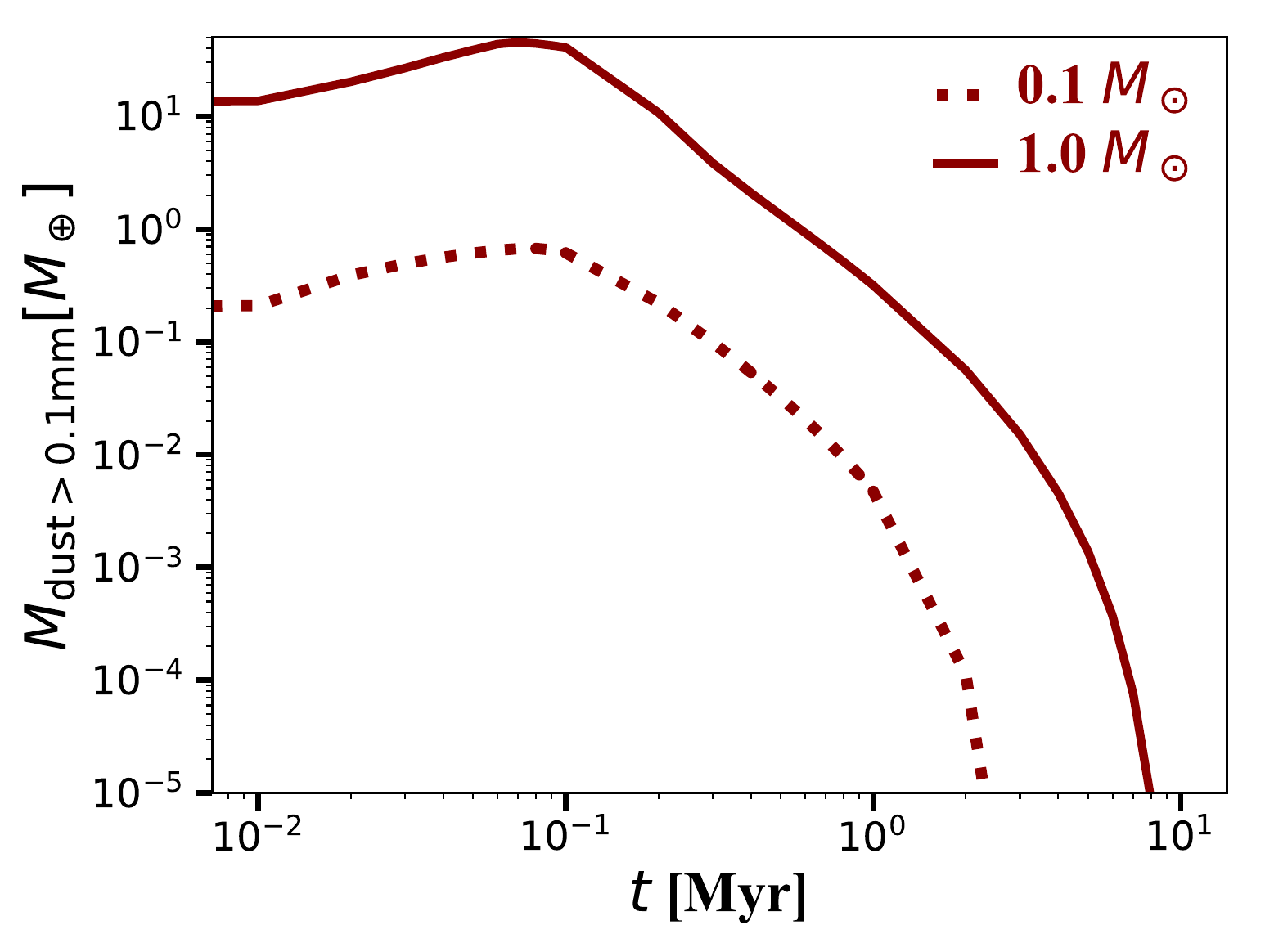}&
    	\includegraphics[width=0.5\columnwidth]{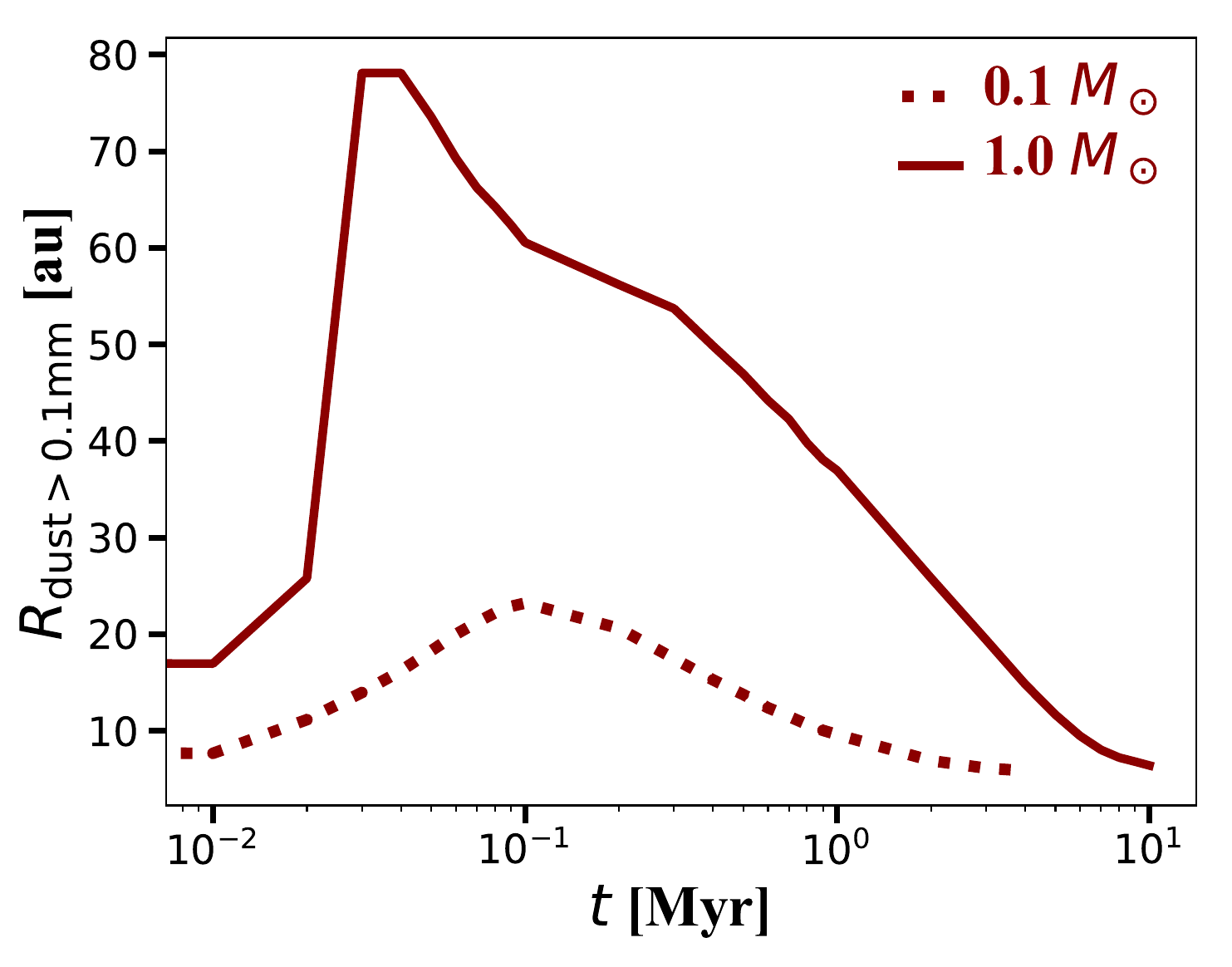}
    \end{tabular}
\caption{Left panel: evolution of the dust mass of grains larger than 0.1\,mm around 0.1\,$M_\odot$ and 1\,$M_\odot$. Right panel: evolution of the radius that encloses 90\% of the dust mass of particles larger than 0.1\,mm around 0.1\,$M_\odot$ and 1\,$M_\odot$. }\label{Mdust_Rdust}
\end{figure}

\noindent where $P$ is the disk gas pressure, $\Sigma_d$ is the dust surface density, and $v_K$ the Keplerian speed. 

Due to the dependencies of $a_{\mathrm{frag}}$ and $a_{\mathrm{drift}}$ on $\Sigma_g$ and $\Sigma_d$ respectively, both barriers correspond to lower grain sizes in the case of the disk around a 0.1\,$M_\odot$. Figure~\ref{dust_evo_variablestar} demonstrates that drift limits the maximum grain size earlier in the disk around the 0.1\,$M_\odot$ star than in the case of a 1\,$M_\odot$ star. Because of the more efficient drift of particles around VLMS, the disk is more depleted of pebbles (here we call pebbles grains larger than 0.1\,mm), making also the dusty disk in the case of 0.1\,$M_\odot$ star smaller with time than in the case of a 1\,$M_\odot$ star.

Figure~\ref{Mdust_Rdust} shows the evolution of the dust mass of grains larger than 0.1\,mm around 0.1\,$M_\odot$ and 1\,$M_\odot$, in addition to the evolution of the radius that encloses 90\% of the dust mass of particles larger than 0.1\,mm. The evolution of $M_{\mathrm{dust}>0.1\rm{mm}}$ shows initially an increase due to dust growth and that the maximum mass of these pebbles is around 50\,$M_\oplus$ for disks around a 1\,$M_\odot$ star and when drift starts to dominate the evolution at around 0.1\,Myr, $M_{\mathrm{dust}>0.1\rm{mm}}$ rapidly decreases with time, in the absence of pressure bumps which can help to form and retain those pebbles in the disk. We note that pressure bumps not only help to retain pebbles, but also to form more of them because inside these regions grain growth is limited by fragmentation, increasing the maximum grain size of particles where pressure bumps are located.

In the case of a 0.1\,$M_\odot$ star, the maximum value of $M_{\mathrm{dust}>0.1\rm{mm}}$ is only about 0.7\,$M_\oplus$ and a faster decrease than in the case of 1\,$M_\odot$ star is seen after 0.1\,Myr. Most of the dust mass of the disks shown in Fig.~\ref{ALMA_VLMS} and in Fig.~\ref{Mdisk_Mstar} (in the context of the $M_{\rm{dust}}-M_\star$ relation) have higher masses than the maximum value of $M_{\mathrm{dust}>0.1\rm{mm}}$ in the case of 0.1\,$M_\odot$ star. It is therefore natural to conclude that the substructures are required (as observed) helping to form and retain pebbles in these disks. 

The evolution of $R_{\mathrm{dust}>0.1\rm{mm}}$ shows how the dusty (pebble) disk increases quickly with time, reaching values near 80\,au in the case of the 1\,$M_\odot$ star and afterwards quickly shrinking with time due to radial drift. In the case of a  0.1\,$M_\odot$ star, the maximum value of $R_{\mathrm{dust}>0.1\rm{mm}}$  is only $\sim24$\,au. Therefore, these disks are expected to be very compact unless pressure bumps are present in the outer disk. 

\section{Gap opening by planets in BDs and VLMS disks}\label{hydro_models}

\begin{figure}
\centering
\centering
    \tabcolsep=0.05cm 
    \begin{tabular}{cc}
        \includegraphics[width=0.5\columnwidth]{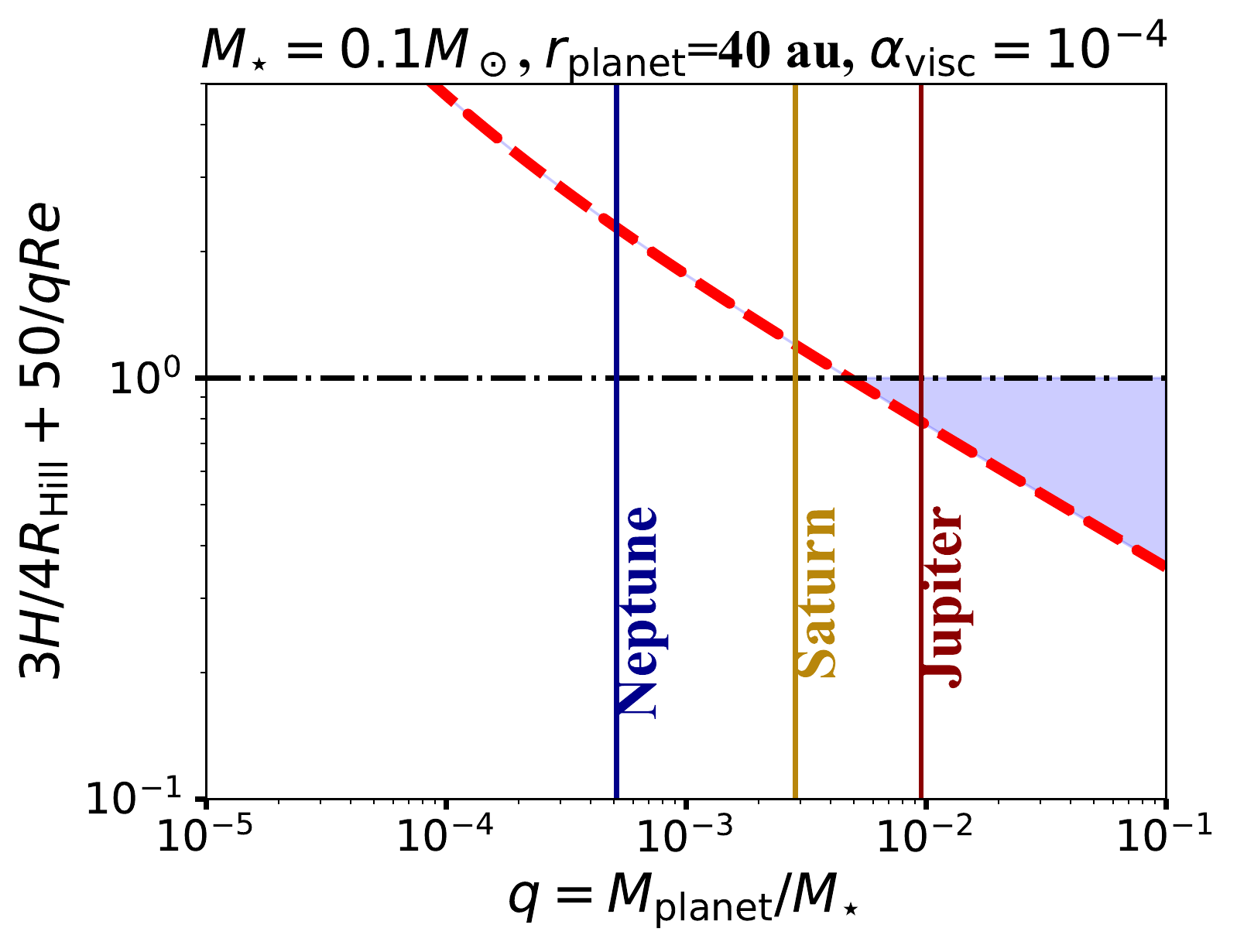}&
    	\includegraphics[width=0.5\columnwidth]{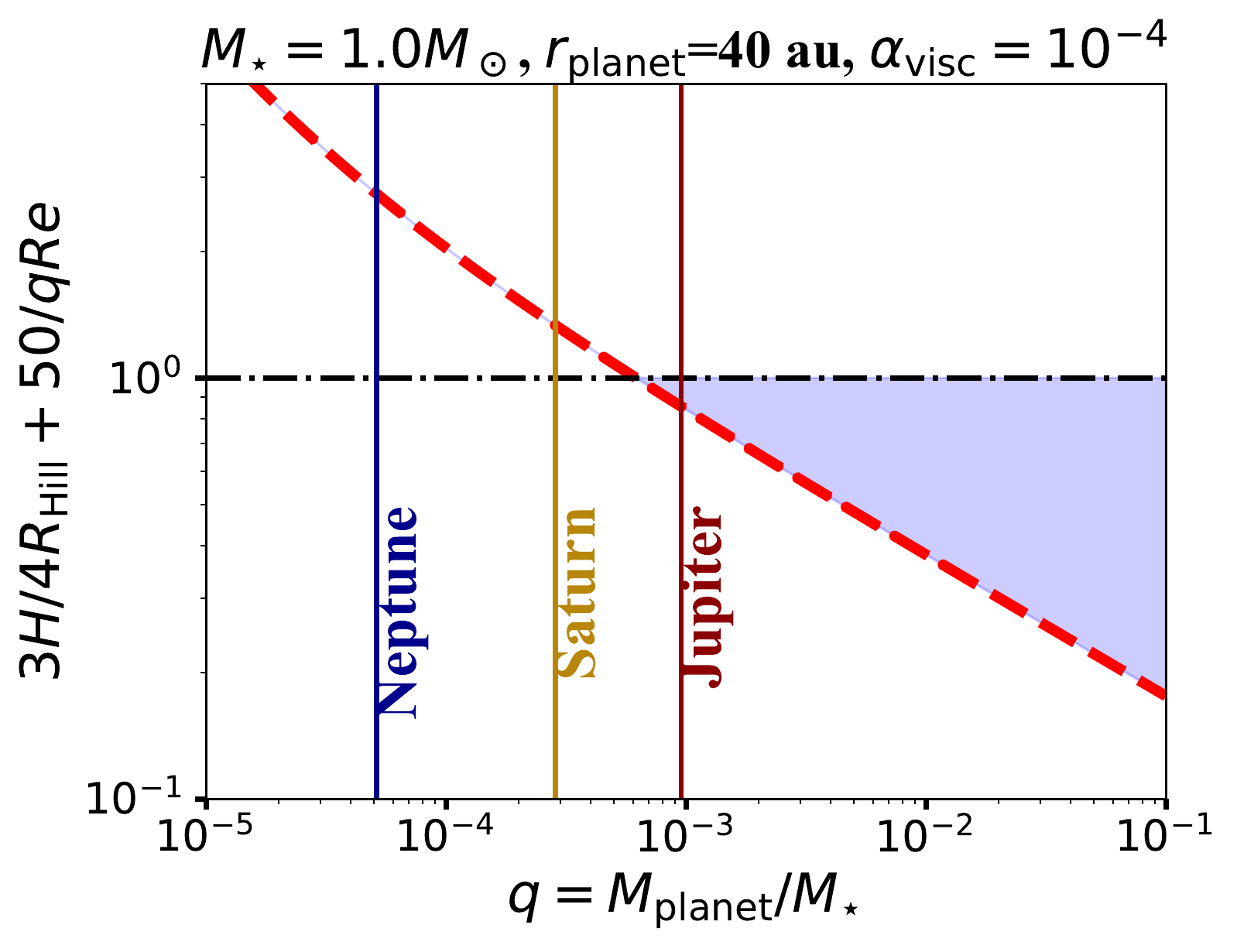}
    \end{tabular}
\caption{Criterion for gap opening (Eq.~\ref{eq:crida}) assuming the stellar parameters of a 0.1\,$M_\odot$ (left) and 1.0\,$M_\odot$ (right) stars at 1\,Myr from the evolutionary tracks in Fig.~\ref{PMS}. In the shaded area the criterion is satisfied.}\label{Crida}
\end{figure}

One of the most common explanations for the presence of pressure bumps in protoplanetady disks is embedded massive planets \citep[e.g.][]{rice2006,zhu2011, gonzalez2012, dipierro2016, rosotti2016, dong2017, zhang2018}. In a disk around a BDs or a VLMS, a higher planet-to-star mass ratio is required to open a gap in the gas surface density and stop the radial drift of particles at the outer edge of the gap. This is not only because drift is higher around BDs and VLMS, but also because these disks are geometrically thicker \citep{mulders2012}, making gap opening more difficult due to the increased pressure forces \citep{sinclair2020}.

The work in \cite{pinilla2017} investigated what is the minimum mass of a planet needed to open a gap in a disk around a BD or a VLMS, using the gap opening criterion by \cite{crida2006}. This criterion considers the disk  viscous torque, the gravitational torque from a planet, and the pressure torque, and it is given by 

\begin{equation}
\frac{3}{4}\frac{H}{R_H}+\frac{50}{q Re}\lesssim 1
\label{eq:crida}
\end{equation}

\noindent where $Re$ is the  Reynolds number at the position of the planet $r_p$, which is equal to $r_p\Omega_p/\nu$ ($\Omega_p$ is the Keplerian frequency at the planet position), with $\nu$ being the disk viscosity ($\nu=\alpha c_s^2/\Omega$, with $c_s$ being the sound speed). In Eq~\ref{eq:crida}, $q$ is the planet-to-star mass ratio, $H$ is the disk scale height and it is equal to $c_s/\Omega$, and $R_H$ is the Hill radius of the planet, i.e. $r_H=r_p(q/3)^{1/3}$. This criterion has been compared by \cite{asensio2021} to the most recent work by \cite{dong2017}, showing that Eq.~\ref{eq:crida} remains valid for different disk and planet parameters.

Assuming the stellar luminosities at 1\,Myr of evolution from the evolutionary tracks in Fig~\ref{PMS} for 0.1 and 1.0\,$M_\odot$ stars, we can calculate the disk scale height at any location assuming that the temperature profile is \citep{kenyon1987}

\begin{equation}
	T(r, L_\star)=T_\star\left(\frac{R_\star}{r}\right)^{1/2} \phi_{\rm{inc}}^{1/4} =\left(\frac{L_\star \phi_{\rm{inc}}}{4\pi \sigma_{\textrm{SB}} r^2}\right)^{\frac{1}{4}}.
  \label{eq:temp}
\end{equation}

Hence, we use Eq.~\ref{eq:crida} to calculate the required planet-to-star mass ratio for a planet located at 40\,au (similar to the typical distances where rings and gaps have been observed in protoplanetary disks \cite{andrews2020}) and results are shown in Fig.~\ref{Crida} for a viscosity of $\alpha=10^{-4}$. The aspect ratio of the disk around a 0.1\,$M_\odot$ star is around 0.15, whereas the typical values of the aspect ratio in a disk around a Sun-type star are $\sim$0.01-0.05. Because of the higher disk thickness, the difference of the required planet-to-star mass ratio is around one order of magnitude higher for the case of a 0.1\,$M_\odot$ star ($q=5\times10^{-3}$), whereas for a 1\,$M_\odot$ we obtain $q=5\times10^{-4}$. In terms of absolute planet mass, the value is similar (near a Saturn mass planet).

\begin{figure}
\centering
\includegraphics[width=1.0\textwidth]{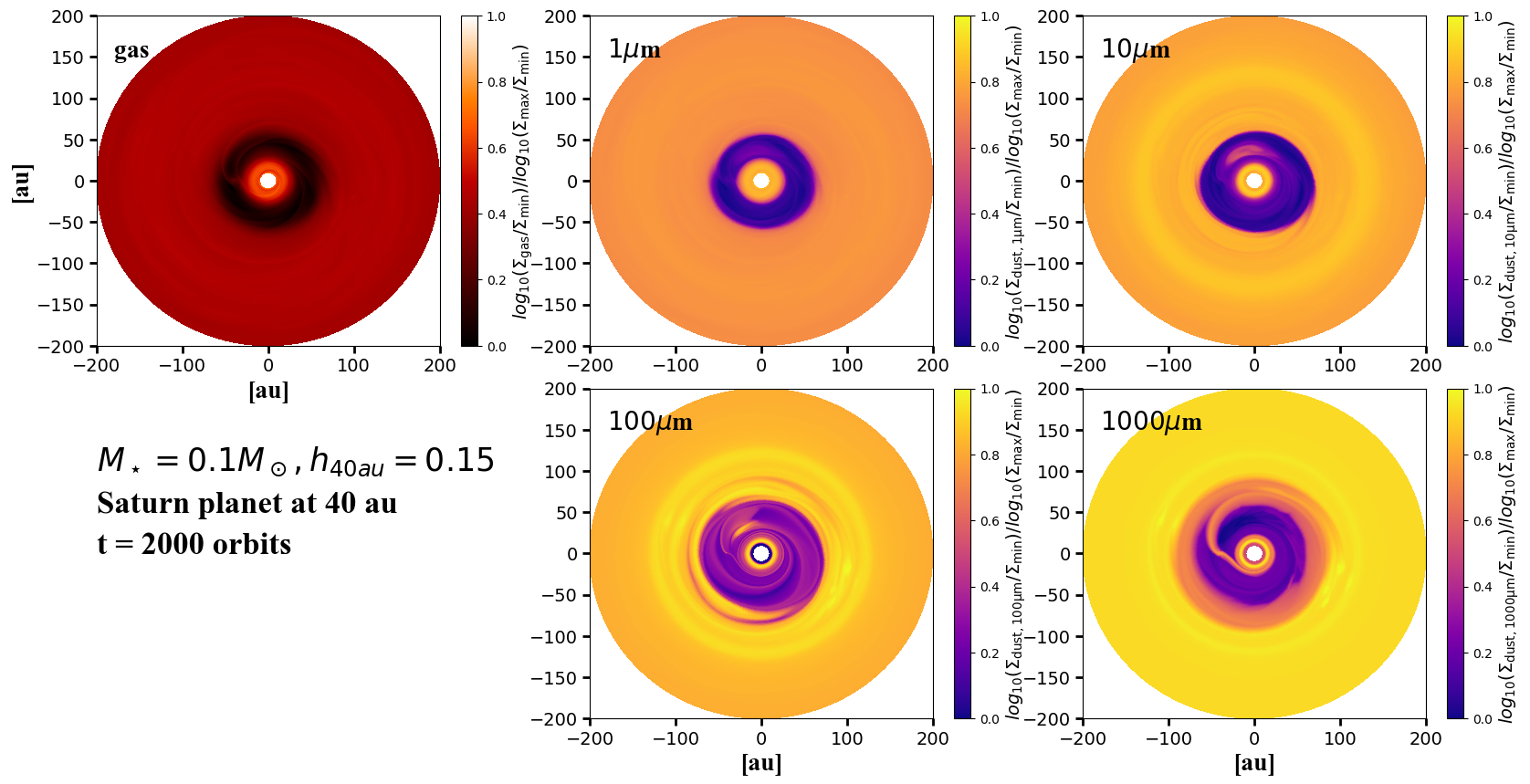}
\caption{Results from hydrodynamical simulations performed with \texttt{FARGO3D} assuming a planet-to-star mass ratio of $5\times10^{-3}$ ($\sim$1.7\,$M_{\rm{Saturn}}$ around a 0.1\,$M_\odot$ star) at 40\,au after 2000 orbits of evolution and assuming $\alpha$-viscosity of $10^{-4}$. The top left panel shows the gas surface density,  the panels from the top middle to the bottom right show the dust surface density of 1, 10, 100, and 1000\,$\mu$m-sized particles, respectively.}\label{hydro_models_fig}
\end{figure}

To quantify how is the distribution of the gas and dust surface density distributions when a planet with a planet-to-star mass ratio of $5\times10^{-3}$ is embedded in a disk around a 0.1\,$M_\odot$ star, we perform hydrodynamical simulations using the publicly available code \texttt{FARGO3D} \citep{benitez2016}. We assume a disk mass of 0.05\,$M_\star$ and the disk extension is from 0.1 to 5\,$r_p$ (4-200\,au). The aspect ratio at the planet location is 0.15, and we assume a flaring index of 0.25 (consistent with the temperature profile from Eq.~\ref{eq:temp}). We include the evolution of 4 dust species with sizes of 1, 10, 100, and 1000\,$\mu$m. These particles are initially distributed as the gas with a dust-to-gas ratio of  0.01. We assume a power-law for the dust grain size distribution, such that $n(a)\propto a^{-3.5}$. The intrinsic volume density of the particles is assumed to be $\rho_s = 1.6~\mathrm{g\,cm}^{-3}$. The planet does not migrate or accrete material during the evolution of these simulations. Planet migration can play an important role for the formation of the small close-in planets that are detected around M-dwarfs, if the observed substructures are particle traps where efficient grain growth happens and where potential small close-in planets can form later while the pressure bump migrates inwards with the giant planet. Upcoming and future observations of close-in planets around M-dwarfs that can provide constraints on the potential migration of these traps are crucial, such as measurements of the core composition (where icy composition would favour migration), and/or the discovery of multiple planets in mean-motion resonances (such as TRAPPIST\,1).

Figure~\ref{hydro_models_fig} shows the gas and dust surface densities after 2000 orbits of evolution. The normalization of each panel is as \cite{zhang2018} to highlight the resulting structures. The results of Fig.~\ref{hydro_models_fig} show that a planet with a planet-to-star mass ratio of $5\times10^{-3}$  clearly opens a gap in the gas surface density and in the distribution of the dust particles of different sizes, characteristic of transition disks. Such gap is similar in shape between the gas and small-sized particles (1-10\,$\mu$m) and it becomes wider for larger particles (100-1000\,$\mu$m); as it has been shown by different works \cite[e.g.][]{pinilla2012, maria2013}. Such difference in the gap shape is testable when comparing  scattered light (tracing small grains) and sub-mm observations (tracing the large grain). This kind of comparison has been done in several transition disks, but it remains to be done for BDs and VLMS because of the current lack of telescopes with the capability of detecting scattered light around BDs and VLMS. In this context, direct imaging with JWST of these objects will provide an important step forward for the understanding of the physical processes ruling the evolution of these disks and the formation of planets around BDs and VLMS. Interestingly, the intermediate-sized particles (10 and 100$\mu$m) show an effective dust trapping around the leading $L_4$ Lagrangian point, similar to recent observations of the disk around the Sun-like star LkCa\,15 \citep{long2022}. Such trapping is less clear in the large particles in this particular case, which depending on dust opacities \citep[e.g.][]{Stadler2022} can lead to the detection of such asymmetry at millimeter-wavelengths.

The type of embedded planets that current models and  observations need to explain the substructures observed in disks around BDs and VLMS (Fig.~\ref{ALMA_VLMS} vs. Fig~\ref{Crida} and~\ref{hydro_models_fig}) represent less than 2\% of the confirmed exoplanets so far  (Fig.~\ref{exoplanets}) around all type of stars. The orbital separation of the giant planets shown in Fig.~\ref{exoplanets} in the VLMS and BD domain is  mostly within the first au, so hot Jupiters. These planets  cannot be invoked to explain the observed disk substructures at tens of au, unless they effectively migrate later in the disk  evolution. The known frequency of giant planets at large radii (that can produce these gaps) is currently smaller than 2\%, and future observations are required to better constrain this occurrence rate.

In the simulations shown in Fig.~\ref{hydro_models_fig}, the presence of an inner disk in the gas and in the dust particles of different sizes remain after several thousands of orbits. However, as it has been shown by \cite{pinilla2021}, such an inner disk is difficult to explain when grain growth is taken into account because the dust particles efficiently grow and drift to the inner disk while the particles from the outer disk are blocked by the gap. In general, there is a current debate how these inner disks survive for long times of evolution in transition disks. 

\section{Summary and Conclusions}\label{summary}

BDs and VLMS are a significant fraction of stars in our galaxy, and they are interesting laboratories to investigate planet formation because the frequency of Earth-sized planets in habitable zones appears to be higher around these stellar objects. In this paper, we  present state-of-the art observations of the dust continuum emission of disks around BDs and VLMS that have been observed with ALMA and show substructures as a motivation to describe one of the main problems during the first steps of planet formation around these objects: the fast radial drift of dust particles. In addition, current ALMA observations of disks around BDs and VLMS suggest that giant planets may originate their structures, but few giant planets have been confirmed around such low-mass stellar objects (Fig.~\ref{exoplanets}), in addition to challenge core accretion models of planet formation in these systems. Our main conclusions are:

\begin{itemize}
    \item The $M_{\rm{dust}}-M_\star$ relation of disks in different star forming regions follow an approximately a power-law relation that steepens in time, and it seems to flatter for disks with substructures. However, the current relation inferred for disks with substructures is obtained with very few data of disks around BDs and VLMS. Dedicated deep observational surveys of BDs and VLMS disks are required to understand the main properties of these objects and to put them in context of the known relations of  $M_{\rm{dust}}-M_\star$ and  $R_{\rm{dust}}-L_{\rm{mm}}$ of their higher-mass stellar counterparts. Such observations will help to understand the main physical mechanisms that rule the evolution of their disks and the effect on the potential planets that may form in these environments. 
    \item A handful of high resolution ALMA observations of disks around BDs and VLMS have been done and showing substructures. Detailed analysis of the origin of the observed substructures in these disks suggest the presence of Saturn- or Jupiter-mass planets. These observations challenge current planet formation models because such massive planets cannot form either by pebble or planetesimal accretion in these low-mass environments, leaving gravitational instability as the unique route to form them. Dedicated surveys for searching for (giant-) planets around BDs and VLMS, such as CARMENES \cite{Quirrenbach2010} and with TESS will soon provide new statistics of exoplanets around these low-mass star objects, helping us to further test models of planet formation in BD and VLMS disks.  
    \item One of the main physical processes for the evolution of the dust in protoplanetary disks is radial drift, which is higher around BDs and VLMS disk during the entire disk evolution. Because of this, with typical disk parameters, the maximum mass of pebbles (particles larger than 0.1\,mm) in these disks is much lower ($<1\,M_\oplus$) than around Sun-like stars ($\sim$50\,$M_\oplus$). In addition, the maximum (pebble) disk size as traced from sub-millimeter observation is much smaller ($\sim$25\,au vs. 80\,au). Both of them, the mass and radial extension of pebbles, decrease sharper with time in BDs and VLMS disks than around Sun-type stars. Multiple pressure bumps of high-amplitude are needed to explain the dust mass (and some of the dust disk sizes) that have been obtained from  ALMA observations of these disks (Fig.~\ref{ALMA_VLMS}).
    \item In a disk around a BDs or a VLMS, a higher planet-to-star mass ratio is required to open a gap in the gas surface density and stop the radial drift of particles at the outer edge of the gap. This is not only because drift is higher, but also because these disks are geometrically thicker, making gap opening more difficult due to the increased pressure forces. The minimum planet-to-star mass ratio to open a clear gap in the gas surface density is around $5\times10^{-3}$ in a disk around a 0.1\,$M_\odot$ star. Hydrodynamical simulations of the gas and dust evolution show that when such a planet is embedded in the disk, inner large gaps characteristic of transition disks are expected. The gaps should be wider when observed in the millimeter wavelength than when observed in scattered light in the near-infrared. With current observational techniques it is not possible to obtained scattered light images of disks around BDs and VLMS, and incoming observations with JWST will unveil the distribution of the small-dust particles in these disks helping to constrain current models of planet formation.   

\end{itemize}

\backmatter

\bmhead{Acknowledgments} 
I am very thankful to several colleagues who inspired me and motivated me to work on the topics presented in this paper, including: Myriam Benisty, Carsten Dominik, Nicolas Kurtovic, Feng Long, Antonella Natta, Luca Ricci, Aleks Scholz, and Leonardo Testi. In addition, thanks to Jun Hashimoto, Feng Long, and Nienke van der Marel for helping me to collect the ALMA observations presented in this paper.  I also thank Matías Gárate for his help implementing the stellar evolution in \texttt{Dustpy}. I acknowledge support provided by the Alexander von Humboldt Foundation in the framework of the Sofja Kovalevskaja Award endowed by the Federal Ministry of Education and Research. 

\section*{Declarations}
The datasets generated during and/or analysed during the current study are available from the corresponding author on reasonable request.

\bibliography{sn-bibliography}% common bib file

\end{document}